\documentclass[12pt]{article}
\usepackage[cp1251]{inputenc}
\usepackage[english]{babel}
\usepackage{graphicx}
\makeatletter
\long\def\@makecaption#1#2{%
  \vskip\abovecaptionskip
  \sbox\@tempboxa{#1. #2}%
  \ifdim \wd\@tempboxa >\hsize
    #1. #2\par
  \else
    \global \@minipagefalse
    \hb@xt@\hsize{\box\@tempboxa\hfil}%
  \fi
 \vskip\belowcaptionskip}
\makeatother

\begin{document}

\begin{titlepage}

\thispagestyle{empty} \vskip 1cm

\title{New Results of Observations of the Only Supernova Remnant in the IC1613 Galaxy.}

\author{Lozinskaya T.A.${}^{1}$,  Podorvanyuk N.Yu.${}^{1}$ }

\maketitle

\centerline{${}^1$\it Sternberg Astronomical Institute, Universitetskii pr. 13, Moscow, 119992 Russia}

\vskip 1cm \centerline{Received 00.00.2008} \vskip 1cm

\end{titlepage}

\begin{abstract}

     {\bf ABSTRACT}
     \medskip

The new results of a study of the kinematics of the supernova
remnant S8 in the IC1613 galaxy are reported. The expansion
velocity of the bright optical nebula is determined based on
observations made with the 6-m telescope of the Special
Astrophysical Observatory of the Russian Academy of Sciences
using MPSF field spectrograph and SCORPIO focal reducer operating
in the scanning Fabry--Perot interferometer mode. An analysis of
21-cm line VLA observations of the galaxy corroborates our earlier
proposed model of a SN exploding inside a cavern surrounded by a
dense shell and S8 colliding with the wall of the HI shell.

\end{abstract}

\section{Introduction}

The only known supernova remnant in the Irr galaxy IC1613 --- the
bright nebula  S8 (Sandage, 1971) --- resides in a giant complex
 of multiple ionized shells (Meaburn et al., 1988; Lozinskaya et
al., 2001; Lozinskaya et al., 2002) in the Northeastern sector of
the galaxy.

Lozinskaya et al. (1998) (hereafter referred to as Paper (I))
showed this remnant to be an x-ray source and observed  S8 at
optical and radio wavelengths. These observations revealed the
peculiar nature of the object, which combines the properties of
an old and a young supernova remnant, i.e., high brightness at
both optical and x-ray ranges. In Paper (I) we proposed a model
where a supernova explodes inside a cavern surrounded by a dense
shell and the expanding SNR collides with the wall of this dense
shell. In this case, bright optical emission is due to the gas
located behind the radiative cooling shock in the dense shell,
whereas bright x-ray emission, to the hot plasma behind the shock
that propagates in the inner tenuous cavern. Rosado et al. (2001)
criticized this model and suggested that half of the remnant in
question is simply unobservable because of dust.

We studied the kinematics of ionized and neutral gas in the
entire complex of ongoing star formation in the galaxy in an
earlier paper (Lozinskaya et al., 2003). In this paper we perform
further analysis of the kinematics of the optical supernova
remnant based on observations made with the 6-m telescope of the
Special Astrophysical Observatory using MPFS spectrograph and
focal reducer SCORPIO operating in the scanning Fabry-Perot
interferometer mode  (Moiseev, 2002), and also search for an
extended dense shell or for a dense HI layer based on the results
of 21-cm line VLA observations.

In the followings sections we report the results of observations
of the supernova remnant made with the MPFS and with SCORPIO
operating in the scanning Fabry--Perot interferometer mode, which
revealed for the first time the expansion of the bright nebula.
The results of VLA observations allowed us to idetify the external
HI shell, thereby supporting the model that  we proposed in Paper
(I). In the Conclusions section we summarize the main conclusions
of this paper.

All velocities mentioned in this paper are heliocentric, and we assume that the distance
to the galaxy is equal to 725--730~kpc (Freedman, 1988).

\section{Structure and Kinematics of the Supernova Remnant According to Observations Made with
the 6-m Telescope using MPFS Spectrograph and Fabry--Perot Interferometer}

\subsection{Observations and Data Reduction}

{\bf MPFS Observations}. In 2001 we performed our observations
with the new version of  MPFS spectrograph, which was an upgraded
version of the instrument employed in Paper I. A description of
the spectrograph can be found at

(\verb*"http://www.sao.ru/hq/lsfvo/devices/mpfs/mpfs_main.html").

Table~1 lists the parameters of two spectrographs. The last row
of the table gives the seeing during the 1996 and 2001
observations.

\begin{table}
\caption{Parameters of the new and old  MPFS.}

\begin{tabular}{|l|l|l|}
\hline
Instrument& Old      & New \\
          &  MPFS       &  MPFS    \\
\hline
Size of the "micropupil"      &$1.3''$ & $0.75''$  \\
\hline
Field of view   &$10.4''$x$20.8''$&$12''$x$11.25''$  \\
\hline
CCD  &1040x1160 px&1024x1024 px  \\
\hline
Wavelength interval     &4000 - 7500 A& 5800 - 7180 A  \\
\hline
Dispersion    &1.6 A/px&  1.35 A/px  \\
\hline
Spectroscopic resolution  &2,5 - 5 A& 3 - 4 A \\
\hline
Seeing &$1.6''$ - $1.7''$& $1.1''$ \\
\hline

\end{tabular}
\end{table}
\bigskip

We reduced our observations using the IDL software developed at
the Special Astrophysical Observatory of the Russian Academy of
Sciences. Primary reduction included bias subtraction,
flatfielding, cosmic-ray hit removal, extraction of individual
spectra from CCD frames, and wavelength calibration of these
spectra using the spectrum of the  He-Ne-Ar filled calibration
lamp. We then subtracted the night-sky spectrum from the
linearized spectra and converted the observed fluxes into
absolute energy scale using spectrophotometric standards observed
at the same zenith distance as IC 1613.

{\bf Interferometric observations} in the H$\alpha$ line were made with SCORPIO focal reducer
and scanning Fabry--Perot interferometer (FPI). A description of the parameters of the
IFP501 interferometer employed and of the reduction technique can be found in the paper
by Lozinskaya et al.~(2003). The size of the field of view was $4.8'$ and the image scale
was $0.56''$/px. The distance between adjacent orders ensured an overlap-free velocity
interval from -2 to -584 km/s; the number of spectral channels was 36, and the width of
a single channel was $\delta\lambda\approx0.36 $\AA, i.e., $\sim16$ km/s in the radial-velocity
scale.

\subsection{Brightness Distribution and Velocity Field in the SNR S8}

Our observations of the supernova remnant with the new version of
MPFS yield HeI (5876 \AA), [OI] (6300 \AA), [OI] (6364 \AA),
H$\alpha$, [NII] (6583 \AA), [SII] (6717+6731\AA), HeI (7065
\AA), and Fe II (7155 \AA) line brightness distribution maps, all
of which have similar structure and agree with the results of
Paper (I). Figure~1 shows the velocity fields based on the
maximum of the contour for each of the  spectral lines mentioned
above as measured with  MPFS spectrograph with the H$\alpha$-line
brightness distribution of the SNR  superimposed.

\subsection{Kinematics of the SNR S8}

In Paper (I) we fitted the H$\alpha$ line  profile by a
superposition of three Gaussians and identified the following
three components: the bright "main" one and two weak components in
the line wings at the level of  10-20\% of maximum intensity. The
latter are the blue and red components shifted by (-150 $\div$ -
423 km/s) and (145 $\div$ 295 km/s), respectively, with respect
to the main component. The main component of the H$\alpha$ line
shows a velocity gradient from -340 km/s in the Western part to
-290 km/s in the Eastern part (see Fig.~6 in Paper (I)).

A reduction of our new  MPFS observations made with superior resolution and seeing compared
to those published in Paper (I) yielded the optimum fit to the H$\alpha$ line
in the form of two bright components -- the red component at about -240 km/s
and the blue one whose velocity differs  from that of the red component by up to -70 km/s.
Figure~2 shows examples of the decomposition of the line profile at different locations
in the nebula according to  MPFS observations.

Our observations made with the scanning FPI are also suggest that the spectral line consists of
two bright components --- the red and the blue --- with almost the same
velocities as those inferred from MPDS data. The velocity of the red component is equal to about
-240 km/s, whereas that of the blue component is offset to within  -100 km/s. Figure~3
illustrates the decomposition of the line profile obtained with the FPI.

In both series of observations the two components have comparable intensities. Note that
we do not discard the weak components in the line wings found in Paper (I). FPI and MPFS
observations record emission at the level of  10--20\% of the maximum intensity in
the velocity intervals from -450 to -150 km/s and from -500 to -100 km/s, respectively.

Figures~4a and 4b show the "position-velocity" (P/V) diagrams for MPFS and FPI observations
of the two bright components of the H$\alpha$ line, which we use to reveal the expansion
of the supernova remnant. We draw three to four P/V diagrams for the directions parallel
to the major axis of the remnant for both observational series and P/V diagrams parallel
to the minor axis based on the results of FPI observations.

These P/V diagrams reveal the expansion of the S8 nebula. The
blue component, which must be produced by the approaching side of
the shell, shows a bended P/V diagram (half of the so-called
"velocity ellipse"), which is typical of an expanding shell. The
expansion velocity of the bright side of the shell as inferred
from this bend is equal to 75 $\pm$ 25 km/s.

The average velocity of the "unshifted" component of ionized gas
emission in the neighborhood of the supernova remnant as inferred
from FPI observations is equal to -240 $\div$ -256 km/s, which is
consistent with earlier observations by Lozinskaya et al. (1998)
and Rosado et al. (2001). According to the results of the
decomposition of the H$\alpha$ line, the second, red component
shows similar velocities. Given that inside the remnant this
component has higher intensity than the  "unshifted" component of
the emission of gas outside the remnant, we assume that it must
result from a superposition of the emission of the receding side
of the shell and that of unperturbed gas along the line of sight.

\bigskip

\section{Extended Outer Shell in the Neighborhood of the SNR}

The results of our detailed analysis of the HI kinematics in the IC 1613 galaxy
based on 21-cm line VLA observations can be found in our earlier papers
(Lozinskaya et al., 2003; Silich et al., 2006).

Figure~5 shows the 21-cm line VLA image of the galaxy region we are interested in
with  H$\alpha$ brightness isophotes superimposed. The arrow points to the supernova remnant.
We computed the HI brightness distribution map by
integrating archive VLA data over 40 channels out  of 127 (in the velocity interval
from -279 to - 178 km/s), because no 21-cm line emission
was recorded from the galaxy in the remaining channels. We computed the
monochromatic H$\alpha$ image by integrating the flux in this line in the spectra
taken with the FPI.

We find the mean HI velocity in the region of the star-forming
complex as inferred from the emission of   "unaccelerated" gas
recorded in all scans to be V$_{HI} = -230\pm 5$ km/s, which is
quite consistent with the estimate of Lake and Skillman (1989)
for this part of the galaxy based on low angular resolution
observations.

It follows from Fig.~5 that the supernova remnant is located at
the outer boundary of the densest HI layer in the galaxy, which
on the South borders shell II (according to the naming convention
of Lozinskaya et al., 2003). The bright shell II is bordered on
the South by another faint neutral shell with the supernova
remnant residing at its inner boundary. Our P/V diagrams shows
signs of the expansion of this neutral shell --- see scan 1
(positions 33 - 63 arcec), scan 2 (90 - 130 arcsec), and scan 3
(60 - 77 arcsec) in Fig~5.

We already mentioned one of these diagrams in our earlier paper
(Lozinskaya et al., 2003) as evidence supporting the model
discussed. The remnant resides at the inner boundary of the shell
and its location in the P/V diagrams in Fig.~5 is indicated by
the arrow. The expansion velocity of the matter in the wall of
the HI shell in the region of the remnant is greater than 10 km/s.

\section{Discussion}

We performed a detailed study of the kinematics of the bright
nebula --- the SNR S8  --- with both MPFS spectrograph and
scanning FPI attached to the 6-m telescope of the Special
Astrophysical Observatory of the Russian Academy of Sciences. We
identified two bright components in the H$\alpha$ line profile
without discarding the weak red and blue components in the wings
of this line found in Paper  (I).

Our "position--velocity" diagrams revealed for the first time the
effect of the expansion of the bright optical supernova remnant S8
with a velocity of  75 $\pm$ 25 km/s.

The inferred expansion velocity suggests that bright optical emission of the remnant is
indeed due to the gas located behind the radiative cooling shock. In Paper I we made this
conclusion based solely on the high brightness of optical radiation.

Note that the expansion velocity inferred from the maximum of the line characterizes the
average velocity of the brightest features; fainter emission is observed over a wide
velocity interval spanning from -500 to -50 km/s. This is a typical situation in all
thoroughly studied bright optical supernova remnants, which is due to the nonuniform
density distribution in the ambient medium. Different cloudlets through which
the radiative-cooling shock propagates differ in density, stages of interaction with the
shock, and in their orientation with respect to the observer.

We used 21-cm line VLA observations to identify the extended
neutral shell at whose inner boundary the supernova remnant
resides. We also revealed the signature of the expansion of this
shell with a velocity of at least  10 km/s. The supernova remnant
is located in the vicinity of the densest region of the HI shell
(see also Lozinskaya et al., 2003.)

The presence of this shell supports the scenario where supernova
exploded inside a cavern surrounded by a dense shell and the
expanding remnant collided with the wall of the shell. We
proposed this model in Paper I to explain the peculiar nature of
the SNR, which combines the properties of a young and old object,
being simultaneously bright both in the optical and x-ray domains.

The collision with this outer shell also helps us to understand
the peculiar morphology of the SNR. The optical nebula exhibits a
well-defined increase of brightness toward the center, which is
typical of plerions. At the same time,  the radio spectrum
($\alpha = -0.6$ according to Paper I, see also Dickel et al.,
1985)) is typical of shell-type SNRs. The elongated
crescent-shaped morphology is due to the fact that at optical
wavelengths we observe only the brightest sector of the
shell-shaped remnant in the dense matter of the outer HI shell.
The offset of the center of the distribution of the velocity of
the bright component of the H$\alpha$ line with respect to the
brightness distribution center, which is evident in Fig.~1,
indicates that the bright sector of the supernova remnant is
tilted with respect to the sky plane.

It goes without saying that this is a model explains the data
provided by the available observations of the SNR S8 . HST
observations are needed to thoroughly understand the nature of
this extremely interesting object.

\section{Conclusions}

We measured for the first time the expansion velocity of the
bright  S8 nebula, which is the only supernova remnant in the
dwarf Irr galaxy IC1613. We revealed an HI shell at whose inner
boundary the SNR resides. This result supports the scenario
proposed in our earlier paper (Lozinskaya et al., 1998) to
explain the peculiar nature of the SNR with the supernova
exploding inside a cavern surrounded by a dense shell and the
supernova remnant then interacting with the wall of the shell.

\emph{Acknowledgments} We are grateful to A.V.Moiseev for
performing observations with the 6-m telescope  and valuable
discussion. This work was supported by the Russian Foundation for
Basic Research (project No.~07--02--00227) and is based on the
observational material obtained with the 6-m telescope of the
Special Astrophysical Observatory of the Russian Academy of
Sciences funded by the Ministry of Science of the Russian
Federation (registration number 01-43). The application for
21-line observations of IC~1613 with NRAO VLA was submitted by
E.Wilcots. The National Radio Astronomy Observatory is a facility
of the National Science Foundation operated under cooperative
agreement by Associated Universities, Inc.

\newpage

FIGURE CAPTIONS

Fig.~1a - Top -- H$\alpha$ image of the S8 supernova remnant superimposed onto the velocity field
in the same line (shown by contours). The flux scale in erg/(s~cm$^{2}~$\AA) is given on the right.
Bottom -- H$\alpha$-line velocity field of the remnant superimposed onto the image taken in the same
line (shown by contours). The velocity scale (km/s) is given on the right.

Fig.~1b - Velocity field of the supernova remnant in various lines superimposed onto
the H$\alpha$ line image  (shown by contours). The velocity scale (km/s) is shown on the right of
each image.

Fig.~2 - Examples of the decomposition of the H$\alpha$ line into
two Gaussians based on MPFS observations. The arrows indicate the
regions of the SNR for which the decomposition is made.

Fig.~3 - Same as Fig.~2, but based on FPI observations.

Fig.~4a - "Position--velocity" diagrams for two components of the
H$\alpha$ line made in the direction parallel to the major axis
of the SNR based on MPFS (left) and FPI (right) observations. The
location of the scans on the image of the nebula is shown at the
top and the velocities of the two line components along these
scans are given by different symbols at the bottom.

Fig.~4b - Same as Fig.~4a, but for the minor axis of the SNR based
on FPI observations.

Fig.~5 - Top -- The 21-cm line image of the galaxy superimposed
on  H$\alpha$ line brightness contours. The thick arrow indicates
the S8 supernova remnant. Arrows 1, 2, and 3 indicate the
locations of the three scans used to construct the 21-cm line P/V
diagrams shown below. The vertical lines indicate the position of
the SNR on the P/V diagrams.

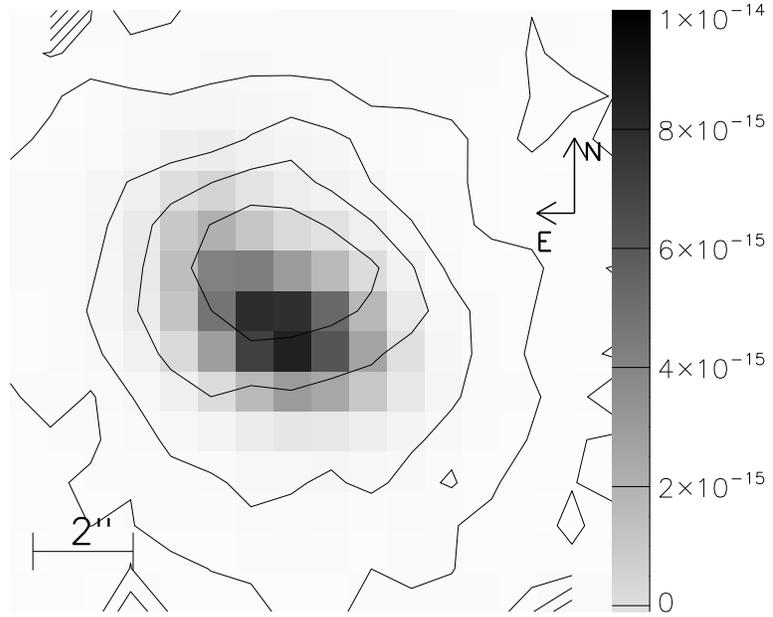
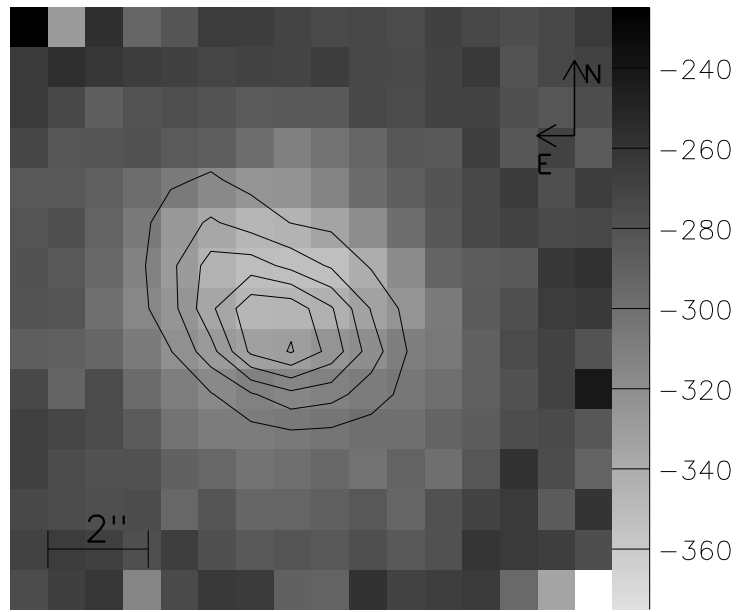
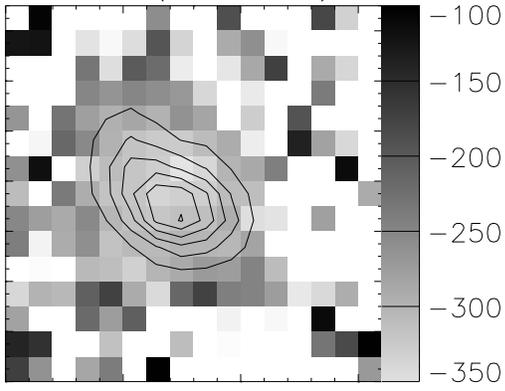
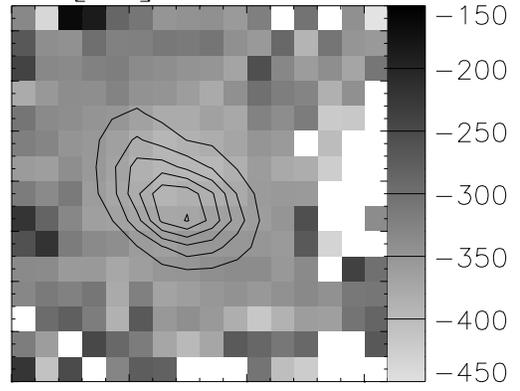
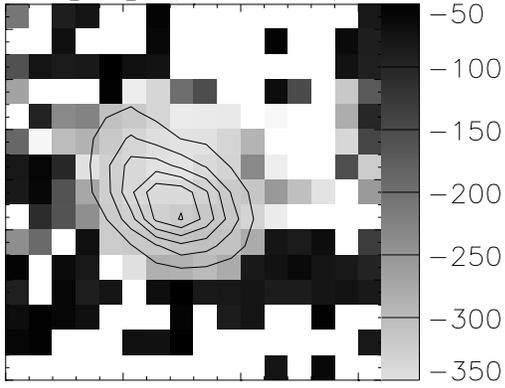
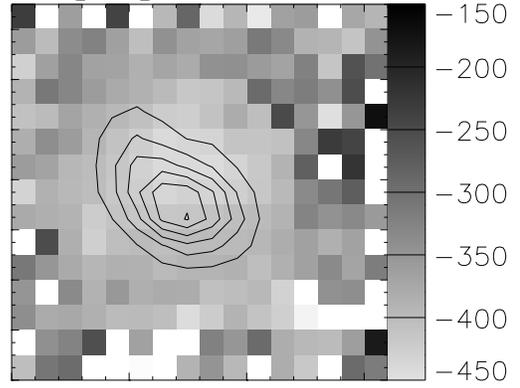
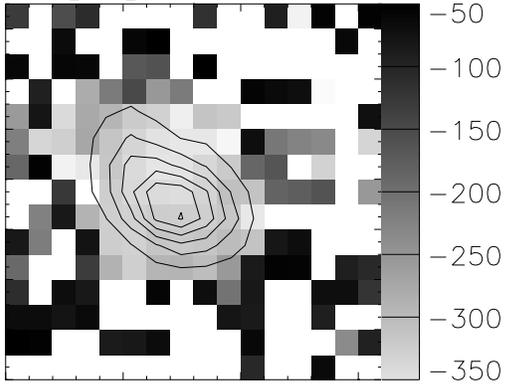
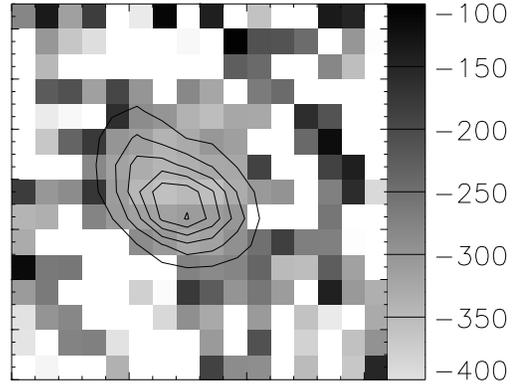
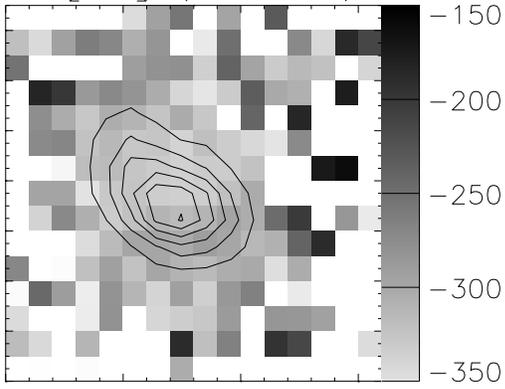
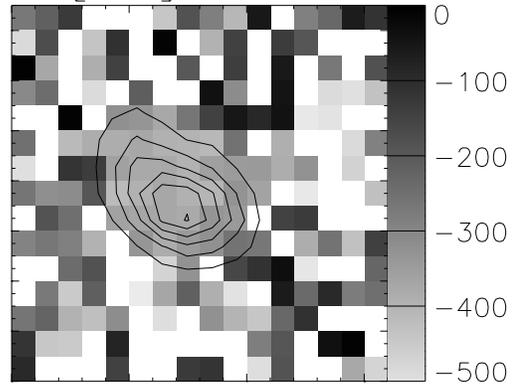
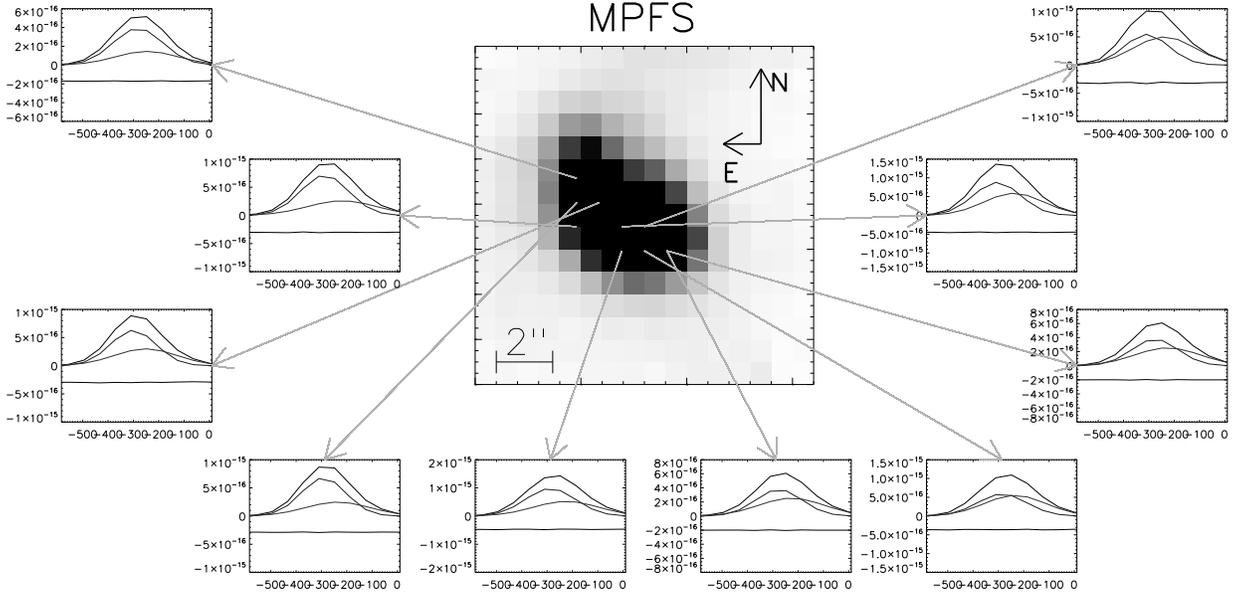
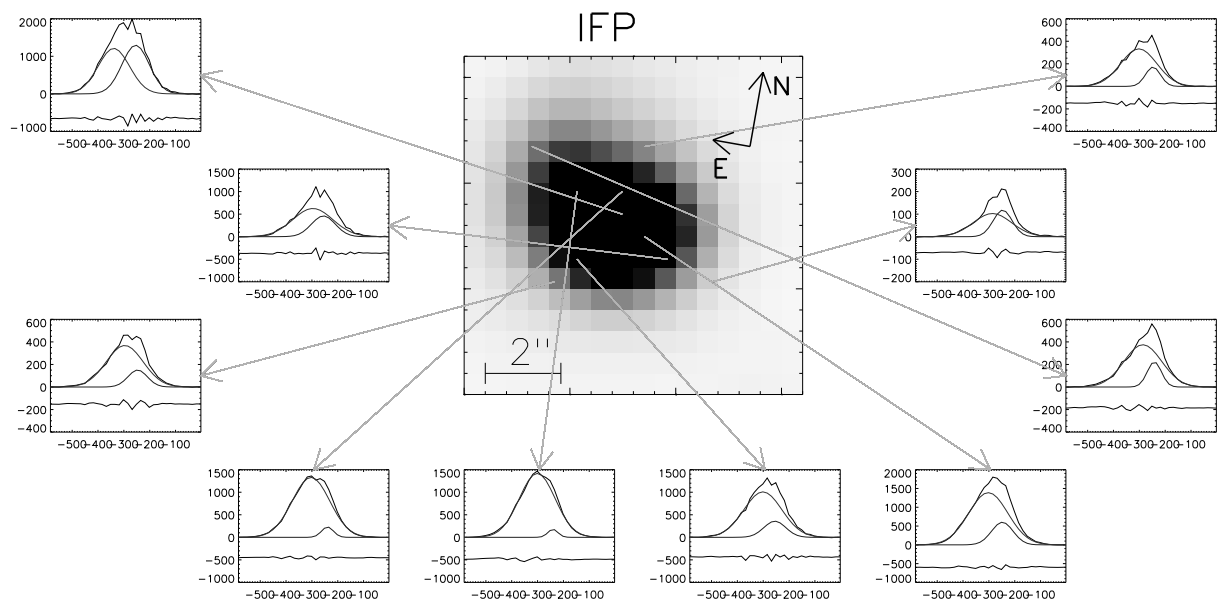
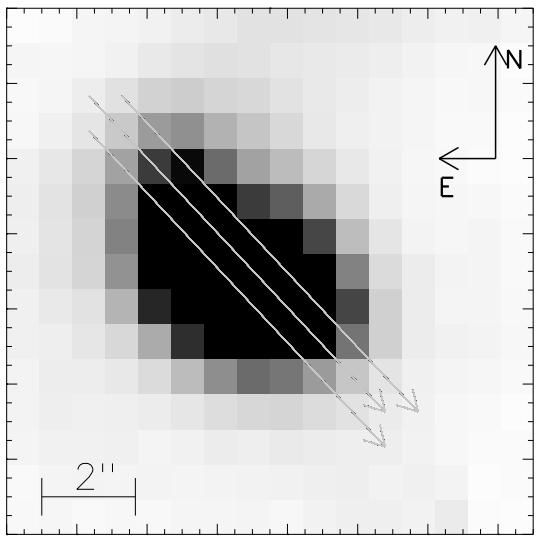
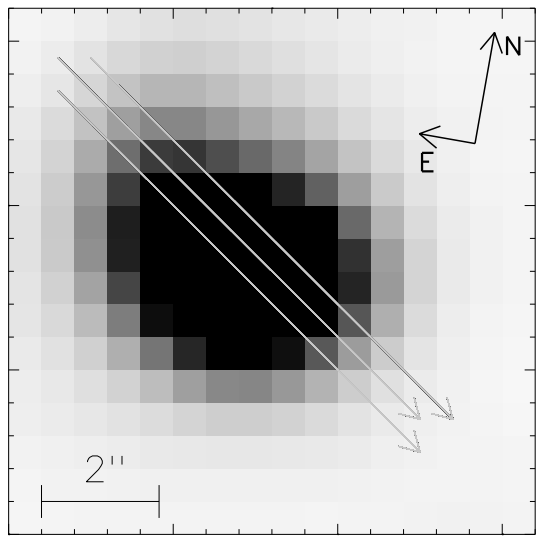
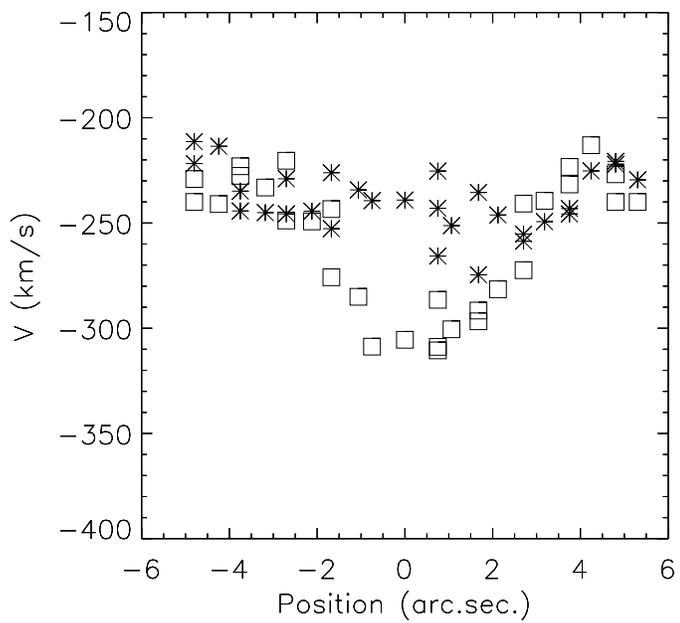
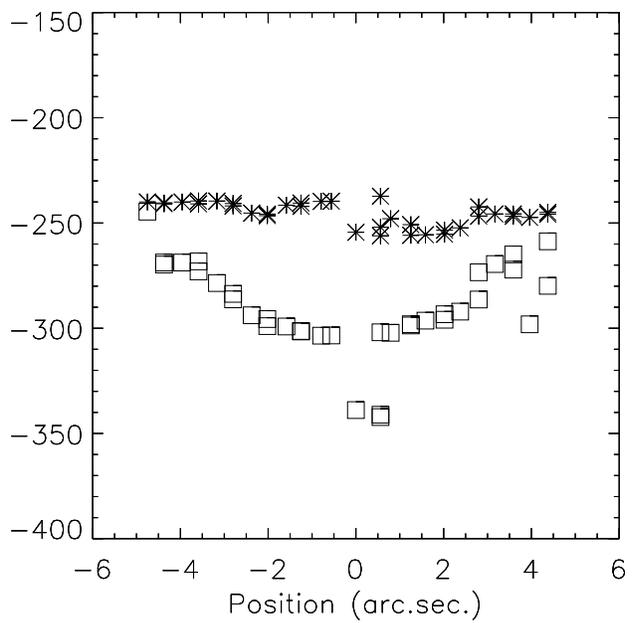
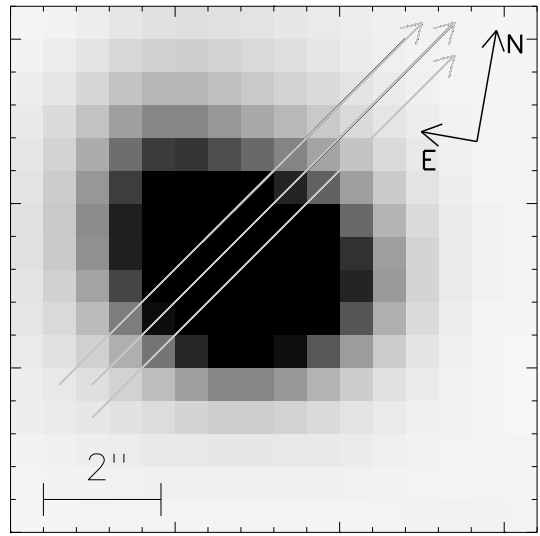
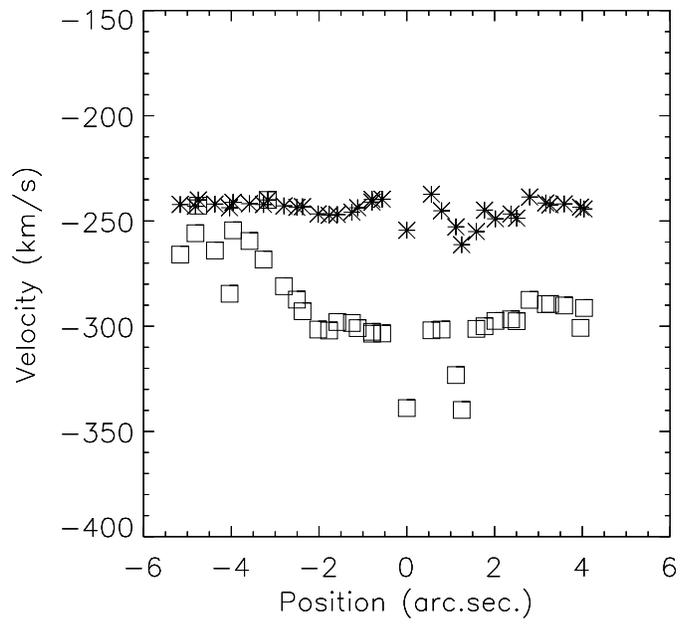
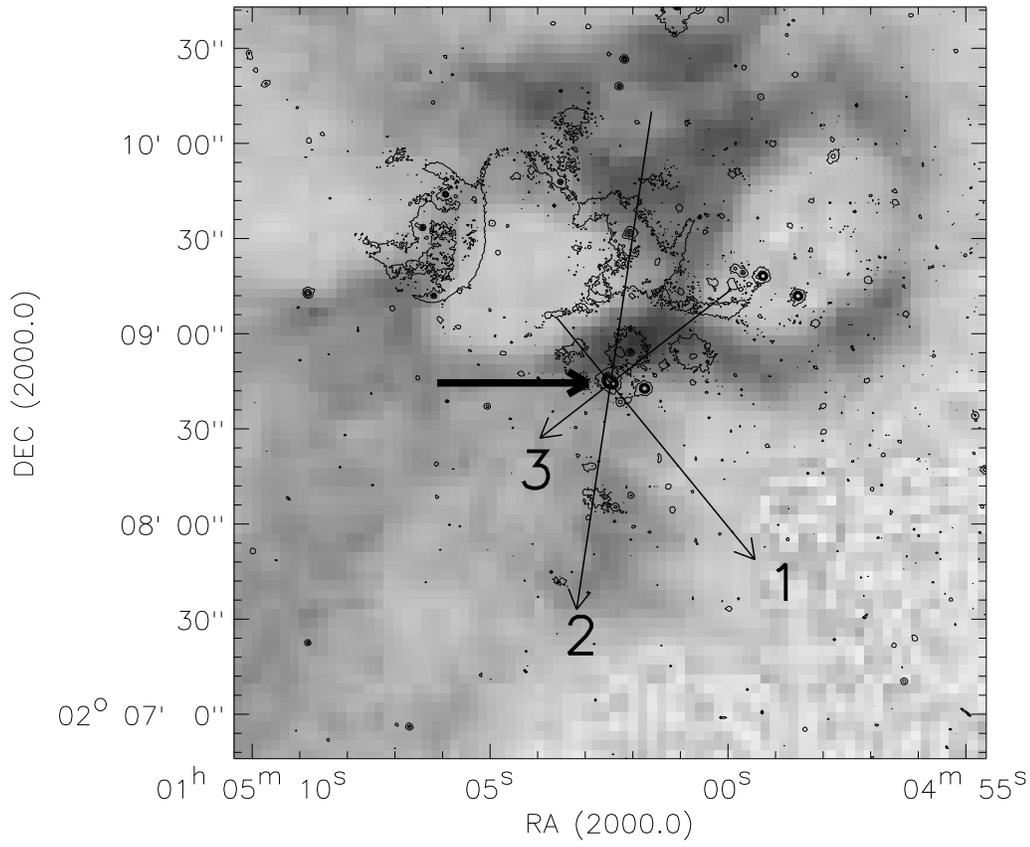
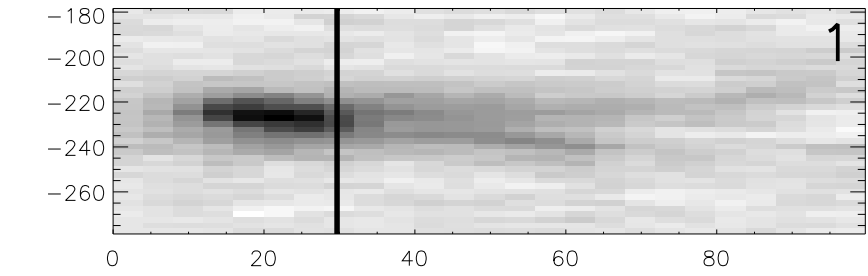
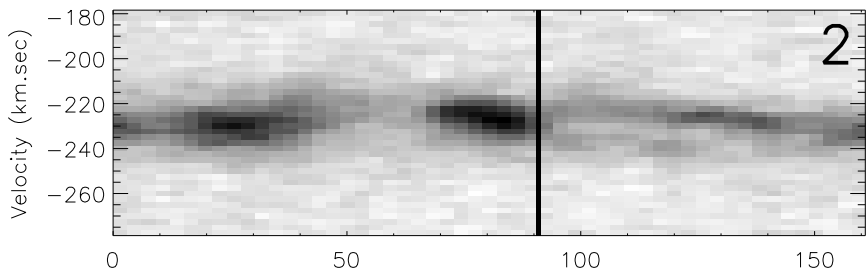
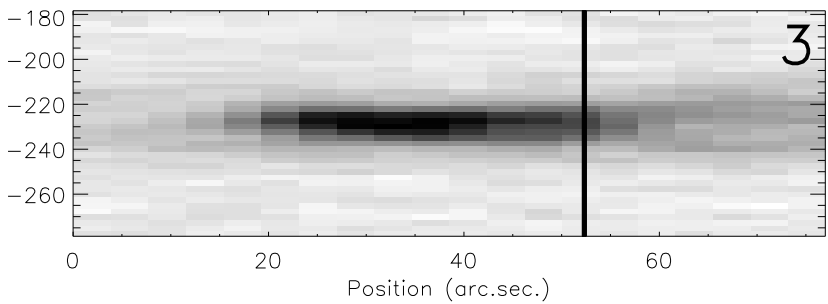
\end{document}